\documentclass[letterpaper,final,conference]{ieeeconf}
\IEEEoverridecommandlockouts
\let\labelindent\relax
\usepackage{enumitem}
\usepackage{amssymb}  % assumes amsmath package installed
\usepackage{cite}
\usepackage{graphics}
\usepackage{epstopdf}
\usepackage[cmex10]{amsmath}
\usepackage{centernot}

\DeclareMathOperator*{\argmin}{arg\,min}
\usepackage{xcolor}
\usepackage{soul}
\usepackage{hyperref}

\newtheorem{theorem}{Theorem}
\newtheorem{lemma}[theorem]{Lemma}
\newtheorem{remark}{Remark}

\newcommand\scalemath[2]{\scalebox{#1}{\mbox{\ensuremath{\displaystyle #2}}}}
\makeatletter
\renewcommand*{\@opargbegintheorem}[3]{\trivlist
  \item[\hskip .22in {\em #1\ #2}] {\em(#3):}\ }
\makeatother
\usepackage{tikz}
\usetikzlibrary{patterns}

\UseRawInputEncoding

\title{\bf Prescribed-Time Safety Design for a Chain of Integrators}%
\author{Imoleayo Abel, Drew Steeves, Miroslav Krsti\'{c}, and Mrdjan Jankovi\'{c}
\thanks{Imoleayo Abel, Drew Steeves, and Miroslav Krsti\'{c} are with the Department of Mechanical and Aerospace Engineering, University of California, San Diego, CA 92093-0411, USA, {\tt\small \{iabel,dsteeves,krstic\}@ucsd.edu}.}%
\thanks{Mrdjan Jankovi\'{c} is with Ford Research and Advanced Engineering, Dearborn, MI 48124 USA, {\tt\small mjankov1@ford.com}}
}

\begin{document}

\maketitle
\thispagestyle{plain}
\pagestyle{plain}

%%%%%%%%%%%%%%%%%%%%%%%%%%%%%%%%%%%%%%%%%%%%%%%%%%%%%%%%%%%%%%%%%%%%%%
\begin{abstract}
Safety in dynamical systems is commonly pursued using control barrier functions (CBFs) which enforce safety-constraints over the entire duration of a system's evolution. We propose a prescribed-time safety (PTSf) design which enforces safety only for a finite time of interest to the user. While traditional CBF designs would keep the system away from the barrier longer than necessary, our PTSf design lets the system reach the barrier by the prescribed time and obey the operator's intent thereafter. %We achieve this by using a time-varying backstepping transformation that utilizes time-dependent gains that grow as time approaches the terminal time. Despite using gains that grow towards infinity, the resulting safe-guarding controller is shown to be uniformly bounded. 
To emphasize the capability of our design for safety constraints with high relative degrees, we focus our exposition on a chain of integrators where the safety condition is defined for the state furthest from the control input. In contrast to existing CBF-based methods for high-relative degree constraints, our approach involves choosing explicitly specified gains (instead of class $\mathcal{K}$ functions), and, with the aid of backstepping, operates in the entirety of the original safe set with no additional restriction on the initial conditions. With Quadratic Programming (QP) being employed in the design, in addition to backstepping and CBFs with a PTSf property, we refer to our design as a QP-backstepping PT-CBF design. For illustration, we include a  simulation for the double-integrator system. 
\end{abstract}

%%%%%%%%%%%%%%%%%%%%%%%%%%%%%%%%%%%%%%%%%%%%%%%%%%%%%%%%%%%%%%%%%%%%%%
\section{Introduction}
\subsection{Hi-rel-deg CBFs}

CBFs have become a popular tool for synthesizing safe controllers for dynamical systems and have been used in a wide range of problem domains: multi-agent robotics \cite{SantilloMulti,MagnusCortesNonsmooth}, robust safety \cite{JANKOVIC2018359,AmesISSCBF,XuRobustness}, automotive systems \cite{JankovicDriverIntent,AmesCruiseControl}, delay systems \cite{TamasCovid,DSCC,JankovicDelay}, and stochastic systems \cite{CLARK2021Stochastic,PrajnaStochastic,CooganStochastic} to mention a few. First defined in \cite{Wieland} and later refined and popularized by the seminal papers \cite{AmesAutomotive,AmesCruiseControl}, CBFs are often employed in a ``safety filter'' framework where they're used for generating safe control overrides for a potentially unsafe nominal controller. In essence, a nominal controller is designed to achieve a desired performance objective, and a CBF-based control override is used whenever the nominal controller is unsafe.

In its initial conception \cite{Wieland,AmesCruiseControl}, CBFs were specified for safety constraints of relative degree one i.e. constraints whose time derivative depend explicitly on the input. The extension of CBFs to constraints with high-relative degree (hi-rel-deg CBFs) was first studied independently in \cite{HsuBipedal,WuSreenathFirstCBFhighRelDeg} with much progress following in~\cite{nguyen2016exponential,XU2018195,xiao2019control,breeden2021high}. In \cite{WuSreenathFirstCBFhighRelDeg}, the extension was limited to the relative degree two case, and in \cite{HsuBipedal} where CBFs of arbitrarily high relative degree was introduced, its usage for a relative degree $r$ case involves choosing $r-1$ bounded, positive definite functions that satisfy additional derivative conditions~\cite[Eq. (26)]{HsuBipedal} -- a requirement that limits the utility of \cite{HsuBipedal} for significantly high relative degree constraints. A similar limitation applies to more recent treatment~\cite{xiao2019control} which requires choosing and tuning $r$ class-$\mathcal{K}$ functions, whose choice determines the subset of the original safe set that is kept forward invariant. Building off \cite{WuSreenathFirstCBFhighRelDeg}, exponential CBFs were reported in \cite{nguyen2016exponential} and allowed the use of simple linear control tools to design CBFs for high relative degree constraints. 
%In particular, the approach reduced the design of high-order CBFs to a pole-placement problem where the poles of a resultant linear system are to be chosen to satisfy relations dependent on the initial condition of the original system for which a CBF is sought. 

%\subsection{The roots of hi-rel-deg CBF designs in ``non-overshooting control''}
\subsection{The ``non-overshooting control'' roots of hi-rel-deg CBFs }

A year before \cite{Wieland}, eight years before \cite{AmesCruiseControl}, and ten years before \cite{nguyen2016exponential}, a design for stabilization to an equilibrium point {\em at} the barrier was introduced, under the name ``non-overshooting control,'' in the 2006 paper \cite{krstic2006nonovershooting} (following its conference version in 2005). This design possesses all the attributes of a safety design with a CBF of a {\em uniform} and high relative degree\footnote{Uniformity of the CBF's relative degree gives the equivalence of a general control-affine system with the strict-feedback class and the convertibility of the safe set, given by the CBF positivity constraint, into  a semi-infinite interval constraint for the first state of the strict-feedback system.} (sans the CBF terminology) with only the QP step absent since, for  stabilization at the barrier, QP is subsumed in the stabilization design (the nominal feedback and the safety-filtered feedback are the same). 

%More than a decade before the advent of CBFs, so-called non-overshooting control topic was gaining interest. It is a special case of a broader problem of control under state constraints, for which control barrier functions and barrier Lyapunov functions are currently popular tools. The special feature of non-overshooting control is that the setpoint for the system is at the boundary of the feasible set. 

The  interest in non-overshooting control in the 1990s came from applications---spacecraft docking, aerial refueling, machining, etc., with no margin for error in downward setpoint regulation. The non-overshooting control problem for linear systems, albeit mostly for zero initial conditions and nonzero setpoints, was solved in~\cite{phillips1988conditions,bement2004construction,schmid2010unified}.  

The paper~\cite{krstic2006nonovershooting} introduced the following two ideas (translated to the current CBF terminology). First, for a system with a hi-rel-deg CBF, a transformation, by backstepping, into a particular target system in the form of a chain of first-order CBF subsystems (resulting in all real poles in the linear case) is performed. Second, in order to ensure that all the CBF ``states'' of this chain begin and remain positive, the positivity of their initial values is ensured by choosing the backstepping gains in accordance with the initial conditions so the entire CBF chain is initialized positively. 

This chain structure and the gain selection of~\cite[Eq. (12), (13)]{krstic2006nonovershooting}, regarded through the lens of pole placement, were independently discovered in the 2016 paper  \cite[Cor. 2]{nguyen2016exponential}.
%\footnote{Compare~\cite[Cor. 2]{nguyen2016exponential} with~\cite[Eq. (12), (13)]{krstic2006nonovershooting}.} 
Likewise, the nonlinear damping choices in the CBF chain in~\cite[Eq. (53), (54)]{krstic2006nonovershooting} was independently proposed in the 2019 paper~\cite{xiao2019control}. Additionally pursued in~\cite{krstic2006nonovershooting}, but not in \cite{nguyen2016exponential,xiao2019control}, was a form of input-to-state safety (ISSf) in the presence of disturbances. This notion, though not explored for hi-rel-deg CBFs, is rigorously conceived in \cite{AmesISSCBF}. 

Inspired by non-overshooting control under disturbances in~\cite{krstic2006nonovershooting}, (i.e. stabilization to an equilibrium at the barrier along with ISSf), mean-square stabilization of stochastic nonlinear systems to an equilibrium at the barrier, along with a guarantee of non-violation of the barrier in the mean sense, is solved in~\cite{WuquanStochasticNonovershooting}.

% Just like in the recent CBF work~\cite{nguyen2016exponential},~\cite{krstic2006nonovershooting} relies on selecting control gains to dominate ratios of the system's initial conditions---in fact,~\cite[Cor. 2]{nguyen2016exponential} and~\cite[Equ. (12), (13)]{krstic2006nonovershooting} are equivalent under certain conditions. This motivates the extension of the techniques in~\cite{krstic2006nonovershooting} to safety-filter design. 

With a nearly negligible QP modification, the stabilizing feedbacks in~\cite{krstic2006nonovershooting} can be used in safety filters. %In fact, we can recast the nonovershooting problem as a safety problem, where our aim is to design a safety-filter to ensure nonovershooting regulation of a system of control barrier functions. The safety-filter is designed specifically to ensure that the control barrier constraint is satisfied for all times. This makes safety-filter design amenable to the backstepping methodology, which provides several advantages. One immediate advantage is that 
Hence, backstepping generates a safety-filter with explicit tuning variables that dictate the exponential approach to the barrier. 
%This feature is uncommon in the safety literature and represents a significant simplification to existing designs that require class-$\mathcal{K}$ functions be designed and tuned for each relative degree.
% Worth remembering is that, in safety-filter design, the goals of ensuring safety and allowing the system to evolve according to a desired nominal trajectory can be conflicting: it was shown in~\cite{reis2020control} that quadratic program (QP)-based methods for safety design may come at the cost of asymptotic convergence to undesirable equilibria. (This does not occur when applying time-varying backstepping for regulation to an equilibrium at the barrier over a finite time interval.) 
%, which generates designs with exponential convergence to the desired equilibrium.

\subsection{Prescribed-time safety (PTSf)}

Recent advances in prescribed-time stabilization (PTS)~\cite{song2017time} have resulted in time-varying backstepping controllers that guarantee settling times independent of initial conditions. Extensions have been developed to stochastic nonlinear systems~\cite{li2021stochastic}, infinite-dimensional systems~\cite{Espitia2018FTSAutomatica,steeves2019bprescribed,steeves2020aprescribed}, and even coupled systems with finite/infinite-dimensional subsystems~\cite{espitia2020sensordelay,steeves2020cprescribed,steeves2020eprescribed}. 
%These designs all share a common feature: convergence to the equilibrium is achieved within a \emph{finite terminal time} that can be a priori prescribed independently of the systems' initial conditions. This type of stabilization, called \emph{prescribed-time stabilization} (PTS), 
PTS is a subset of both the finite-time~\cite{haimo1986finite} and fixed-time~\cite{polyakov2015finite} notions (i.e., stronger than both).

The success in achieving stabilization in prescribed time, independent of initial conditions, raises the question of pursuing the safety counterpart of the same notion. The ``translation'' from stability to safety may be a bit counterintuitive: while PTS guarantees that the state {\em reaches the equilibrium \underline{no later}} than a prescribed time $T$, PTSf guarantees that the state {\em cannot reach the barrier \underline{sooner}} than a prescribed time $T$. 

In what context is such a PTSf property useful? First, it should be noted that PTSf is ``less safe'' than exponential safety (ESf). Less safe is useful by not being needlessly conservative. If the desired operation is in the barrier's proximity, and especially if the desired operation is beyond the barrier, where it is safe to be after time $T$, PTSf offers obvious performance (or alertness) advantages over ESf. It is even known in the automotive area that ``too safe'' may mean very unsafe: a follower vehicle that keeps a large distance `invites' vehicles from other lanes to cut in~\cite{8569979}.

We distinguish our notion of PTSf from existing notions of limited duration safety~\cite{MagnusLimitedDurationCBF}, and periodic safety using fixed-time CBFs~\cite{PanagouAmesFxTCBF}. Specifically,~\cite{PanagouAmesFxTCBF} introduced the notion of periodic safety where the objective is to keep a system safe for \emph{all times} while enforcing that it periodically (with time period $T$) visits a goal set \emph{inside} the safe set. In~\cite{MagnusLimitedDurationCBF} the notion of limited duration safety was studied, and like PTSf it implies that a system is kept safe only for a limited duration $T$. While \cite{MagnusLimitedDurationCBF} restricts the set of initial conditions---a set that shrinks as $T$ increases~\cite[Rk. 2]{MagnusLimitedDurationCBF}---to be a strict subset of the safe set~\cite[Eq. (3)]{MagnusLimitedDurationCBF}, our notion of PTSf places no restriction on the initial conditions of the system.

%Whether maintaining safety exponentially in time or more slowly, existing QP-based safety filters, which override the nominal control as the state dangerously approaches the barrier, could be unnecessarily conservative. Even if the system needs to perform just a {\em finite-time} maneuver, the safety filters maintain a distance from the barrier for infinite time, as if the system will operate (and exist) forever. Hence, applying time-varying backstepping to safety is of theoretical and practical interest.

Finally, two distinct features of the time-varying backstepping technique make it quite attractive for use in safety-filter design. The first is that, compared to ESf designs, the PTSf filters designed with time-varying backstepping do not exhibit large transients when the safety filter overrides the nominal controller. This is not the case for ESf filters  with rapid decay rates: the so-called ``peaking" phenomenon~\cite{kimura1981new,sussmann1991peaking,khalil1992semiglobal} is exhibited, which causes some of the states to become very large near the initialization time, before rapidly converging to the equilibrium. This behavior can cause large state-derivatives, which, e.g., is undesirable in vehicle systems where maneuvers causing large acceleration and ``jerk" can be dangerous. PTSf safety-filter designs avoid peaking by using small gains near initialization time that only grow large as the state grows ``small".

The second feature making time-varying backstepping attractive is the behavior of the convergence it achieves near the terminal time. PTSf achieves convergence with ``infinitely-soft" landing, that is, the state and \emph{all} of its derivatives converge to the equilibrium by the terminal time. This feature occurring in finite time is unique to PTSf, and is desirable because it can ensure, e.g., ``jerk-free" safety maneuvers by the terminal time.

% \subsection{Contributions}
% {\color{blue}Recap. advantages here}

% {\color{red} I rarely see this done in conference papers, especially when the introduction enumerates the benefits of the work in the paper}

%%%%%%%%%%%%%%%%%%%%%%%%%%%%%%%%%%%%%%%%%%%%%%%%%%%%%%%%%%%%%%%%%%%%%%
\section{Problem Description and Preliminaries}
\subsection{Problem Description}
We study systems in the chain-of-integrator form 
\begin{equation}
\begin{aligned}
\dot x_i(t) &= x_{i+1}(t),\quad i=1,\dots,n-1, \\%+ \varphi_i(\underline x_i)\\
\dot x_n(t) &= u(t), \\%+ \varphi_n(\underline{x}_n)\\
y(t) &= x_1(t),\quad\quad t\geq t_0,
\end{aligned}\label{sys}
\end{equation}
with relative degree $n$, where $t_0\geq 0$ is the initialization time, $x(t)=[x_1(t),\dots,x_n(t)]^\top\in\mathbb{R}^n$ is the state with initial condition $x(t_0)$, and $u(t)\in\mathbb{R}$ is the control input. The control objective is to enforce ``safety'', defined here as the non-positivity of the output $y(t)$ over the finite time horizon $[t_0,t_0+T)$, where $T$ is a terminal time that can be a priori prescribed. % As is the case in many safety-critical systems, the control input will be filtered through a so-called safety-filter which overrides a nominal control input $u_{\text{nom}}$ whenever it violates conditions that lead to $y(t)>0$. Distinct to our safety-filter design is that it only enforces safety for times $t\in[t_0,t_0+T)$, and our methodology generates an explicit safety-filter expression which depends on $x(t_0)$ and parameters that can be readily tuned for system performance objectives.

\subsection{Preliminaries}
We denote by `$\mathbb{N}$' the set of natural numbers excluding 0. Our PTSf designs will be generated by the following ``blow-up'' function:
\begin{align}\label{eq:mu_def}
\mu_m(t-t_0,T) &= \frac{1}{\nu^m(t-t_0,T)},\quad t\in[t_0,t_0+T)
\end{align}
for $m\in\mathbb{N}_{\geq 2}$ and the \emph{terminal time} $T>0$, where
\begin{align}\label{eq:nu_def}
\nu(t-t_0,T) &:= \frac{T+t_0-t}{T}
\end{align}
decays linearly from one to zero by the terminal time. We denote by $m^{\overline{k}}$ the \emph{rising factorial} for $m,k\in\mathbb{N}$, that is,
\begin{align}
m^{\overline{k}}&:=m(m+1)\cdots(m+k-1);
\end{align}
the derivatives of $\mu_m$ are 
\begin{align}\label{eq:mu_derivative}
\mu_m^{(i)}(t-t_0,T)=\frac{m^{\overline{i}}}{T^i}\mu_{m+i}(t-t_0,T).
\end{align}
For the rest of this paper, we shall use $\mu_m$ and $\mu_m(t)$ to denote $\mu_m(t-t_0,T)$ for brevity when there is no confusion.\\
We denote my $\mathcal{P}^n(x)$ an $n$th-order polynomial in $x$.

\section{$N$th-Order Chain-of-Integrators Design}\label{sec:nthorder}
% \begin{equation}
% \begin{aligned}
% \dot x_i(t) &= x_{i+1}(t), \quad i=1,\dots,n-1,\\%+ \varphi_i(\underline x_i),\quad i=1,\dots,n-1\\
% \dot x_n(t) &= u(t), \\%+ \varphi_n(\underline{x}_n)\\
% y(t) &= x_1(t),\quad\quad t\geq t_0.
% \end{aligned}\label{sysnodelay}
% \end{equation}
% We aim to design a safety-filter such that $y(t)\leq0$ uniformly over the finite time horizon $[t_0,t_0+T)$; that is, we wish to only enforce safety while it may be needed. 

% We begin our design by performing a time-varying backstepping transformation defined as follows
% \begin{equation}
% \begin{aligned}
% h_1 &:= -x_1\\
% h_2 &:= -x_2 + c_1\mu_2 h_1 
% \end{aligned}\label{sec3:htransform}
% \end{equation}
% where $c_1$ is a design parameter to be determined. We call the transformed states $h_i$'s barrier functions to connote the desire to keep their values positive, provided that the initial values $h_i(t_0)$ are positive as is typical with CBFs. The positivity of $h_1(t_0)$ follows from requiring that the system is initially safe i.e. $x_1(t_0)<0$. For $h_2(t_0)$, we achieve positivity by choosing $c_1>\max\left\{0,-\frac{x_2(t_0)}{x_1(t_0)}\right\}$ where we have used $\mu_2(t_0-t_0,T)=1$ in \eqref{sec3:htransform}. The importance of the restriction of $c_1$ to positive values will become apparent from the dynamics of the target $h$-system. Under the transformation \eqref{sec3:htransform} and the choice of control

We begin by performing a time-varying backstepping transformation defined as follows, for $t\in[t_0,t_0+T)$:
\begin{align}
h_i &:= -x_i+\alpha_{i-1}(\underline{x}_{i-1},t),\quad i=1,\dots,n,\label{hieqn}\\
\alpha_0(\underline{x}_{0},t) &\;\equiv 0,\\
\alpha_i(\underline{x}_i,t) &:= c_{i}\mu_2 h_i+\frac{d}{dt}\alpha_{i-1}(\underline{x}_{i-1},t),\ \  i=1,\dots,n,\label{alphaieqn}
\end{align}
where $c_i$, $i=1,\dots,n$ are design parameters to be determined. We call the transformed states $h_i$, $i=1,\dots,n$ barrier functions to connote the desire to keep their values positive, provided that the initial values $h_i(t_0)$ are positive as is typical with CBFs. For the control input, we use the safety-filter
%with the safety-filter
\begin{align}
u &= \begin{cases}
\min\left\{u_{\text{nom}},\ \alpha_n(\underline{x}_n,t)\right\},&\text{if }t_0\leq t<t_0+T,\\
u_{\text{nom}}g(t,x_1(t_0+T)),&\text{if }t\geq t_0+T,
\end{cases}\label{ueqn}
\end{align}
where $u_{\text{nom}}$ is a potentially unsafe nominal controller and $g$ is the ``ramp" function
\begin{align}
&g(t,x_1(t_0+T))\label{eq:rampfunction}\\
&:=\begin{cases}1-\nu^m(t-t_0-T,\bar{T}),&\text{if } x_1(t_0+T)=0, \\ & t_0+T\leq t\leq t_0+T+\bar{T},\\
1,& \text{otherwise},
\end{cases}\nonumber
\end{align}
where $m\in\mathbb{N}$ and $\bar{T}>0$ are design parameters. The role of $g$ in the product $u_{\mathrm{nom}}g(t,x_1(t_0+T))$ in~\eqref{ueqn} is to ensure that the control law is continuous at $t=t_0+T$, since we will show that the feedback law $\alpha_n(\underline{x}_n,t_0+T)=0$ (cf. Section~\ref{sec:pf_main}). As defined, the continuity of the controller \eqref{ueqn} follows from the continuity of $u_{\text{nom}}$.%Hence,~\eqref{ueqn} overrides a potentially ``dangerous" nominal controller only for $T$ time units.

Strictly speaking, the safety-filter %$\min\left\{u_{\text{nom}},\ \alpha_n(\underline{x}_n,t)\right\}$ from 
\eqref{ueqn} during times $t_0\leq t< t_0+T$ is the solution of the QP problem
\begin{equation}
\begin{aligned}
    u= &\argmin_{v\in\mathbb{R}}\ |v - u_{\text{nom}}|^2\\
   &\text{subject to}\ \ v \leq \alpha_n
\end{aligned}
\end{equation}
where constraint $v\leq\alpha_n$ is equivalent to $\dot h_n+c_n\mu_2h_n\geq0$ under input $v$. Therefore, we refer to our design as a QP-backstepping PT-CBF design.

With the safety-filter \eqref{ueqn}, the CBFs \eqref{hieqn} satisfy
\begin{align}
\dot h_i &= -c_i\mu_2 h_i + h_{i+1},\quad i=1,\dots,n-1,\label{hidoteqn}\\
\dot h_n &\geq -c_n\mu_2 h_n,\label{hndoteqn}
\end{align}
for $t\in[t_0,t_0+T)$. We can now state our main result.
\begin{theorem}\label{thm:main}
If the system~\eqref{sys} is initially safe, that is, $y(t_0)<0$, then the controller \eqref{hieqn}--\eqref{ueqn} ensures that $y(t)<0$ for all $t\in[t_0,t_0+T)$ for the \emph{initial} control gains
\begin{equation}
c_i > \max\left\{0,\underline{c}_i\right\},\quad i=1,\dots,n-1,\label{cieqn}
\end{equation}
where
\begin{equation}
\begin{aligned}
\underline{c}_i &= \frac{x_{i+1}(t_0)-\frac{d}{dt}\alpha_{i-1}(\underline{x}_{i-1}(t_0),t_0)}{\alpha_{i-1}(\underline{x}_{i-1}(t_0),t_0)-x_i(t_0)},\label{ciueqn}
\end{aligned}
\end{equation}
and $c_n\geq 0$. Moreover, the control law~\eqref{ueqn} is uniformly bounded provided that $u_{\mathrm{nom}}$ is continuous in the interval $[t_0,t_0+T]$.
\end{theorem}
\begin{remark}
While not characterized in Theorem~\ref{thm:main}, if the safety filter overrides the nominal controller over the time interval $[t_0+\bar{t},t_0+T)$ for some $\bar{t}<T$, then the convergence of the CBFs to zero with be ``infinitely-soft": that is, all of the derivatives $\frac{d^kh_i(t)}{dt^k}$, $k\in\mathbb{N}$, will also converge to zero by the terminal time $t_0+T$. This also holds true for the $x$-system states. %This is advantageous in many applications, e.g., when performing overriding safety maneuvers for vehicles, this feature ensures the maneuver to be ``jerk-free" at the terminal time. 
See Section~\ref{sec:pf_main}, and in particular,~\eqref{eq:limit_der_h_n} and~\eqref{eq:limit_der_h_i} for the mathematical treatment of this ``infinitely-soft" convergence.
\end{remark}

%We now pursue a proof of Theorem~\ref{thm:main}.
\section{Proof of Theorem~\ref{thm:main}}\label{sec:pf_main}

The structure of our proof comes in two parts: one to establish non-positivity of $y(t)$ for $t\in[t_0,t_0+T)$; and another to establish uniform boundedness of the control law which filters the nominal controller to enforce safety. % While these two parts are intimately connected, each part requires different treatments. To ensure positivity of $h_1(t)$ for $t\in[t_0,t_0+T)$, it is enough to invoke the control barrier \emph{constraint}~\eqref{hndoteqn} and our choice of control gains~\eqref{cieqn}. On the other hand, due to the possible switches in control input between $u_{\mathrm{nom}}$ and $\alpha_n$ in~\eqref{ueqn}, one must take care to ensure uniform boundedness of the control input $u=\alpha_n$, whose gains increase with time according to~\eqref{alphaieqn} \emph{even when} $u=u_{\mathrm{nom}}$. 
To this end, we first present the following commutativity property of the ``blow-up" function~\eqref{eq:mu_def} which we will leverage to show controller boundedness. To simplify our presentation, we take $t_0=0$ henceforth.
\begin{lemma}\label{lemma:blow_ups_commute}
For $m\in\mathbb{N}_{\geq1}$ and $0\leq\bar{t}\leq t<T$, the ``blow-up" function~\eqref{eq:mu_def} satisfies
\begin{align}
    \mu_{m}(t,T)=\mu_m(\bar{t},T)\mu_m(t-\bar{t},T-\bar{t}).
\end{align}
\end{lemma}

\emph{Proof:} Omitted due to space limitation.

% \begin{proof}
% It follows directly from the definition~\eqref{eq:mu_def}:
% \begin{align}
%     \mu_m(t,T)&:=\frac{1}{\left(1-\frac{t}{T}\right)^m}\nonumber\\
%     &\;=\frac{1}{\left(1-\frac{\bar{t}}{T}\right)^m}\frac{1}{\left(\frac{T-t}{T-\bar{t}}\right)^m}\nonumber\\
%     &\;=\mu_m(\bar{t},T)\mu_m(t-\bar{t},T-\bar{t}).
% \end{align}
% \end{proof}

\noindent To demonstrate controller uniform boundedness, we must leverage the fact that the feedback law invokes PTSf whose convergence dominates the rate of divergence of the time-varying control gains in~\eqref{alphaieqn}. To accomplish this, we characterize the following property of the closed-loop system.

\begin{lemma}\label{lemma:exp_soft_landing}
For $c>0$, the $i$th derivative of the function
\begin{align}
    \xi(t):=e^{-cT\left(\mu_1(t,T)-1\right)}
\end{align}
satisfies
\begin{align}\label{eq:soft_landing_limit}
    \lim_{t\rightarrow T^-} \frac{d^i\xi(t)}{dt^i}=\lim_{t\rightarrow T^-}\mathcal{P}^{2i}\left(\mu_1(t,T)\right)\xi(t)=0,\ i\in\mathbb{N}.
\end{align}
\end{lemma}

\emph{Proof:} Follows by induction and repeated application of l'H\^{o}pital's rule. Details to be included in a journal version.

% \begin{proof}
% We compute the first derivative according to~\eqref{eq:mu_derivative}:
% \begin{align}
%     \frac{d\xi(t)}{dt}&=-2c\mu_2(t,T)e^{-cT\left(\mu_1(t,T)-1\right)}\label{eq:xi_1st_der}
% \end{align}
% An application of l'H\^{o}pital's rule to~\eqref{eq:xi_1st_der} twice verifies~\eqref{eq:soft_landing_limit} for $i=1$, since
% \begin{align}
%     \lim_{t\rightarrow T^-}\frac{d\xi(t)}{dt}&=-2ce^{cT}\lim_{t\rightarrow T^-}\mu_2(t,T)e^{-cT\mu_1(t,T)}\nonumber\\
%     &=-2ce^{cT}\lim_{\tau\rightarrow +\infty}\frac{\tau^2}{e^{cT\tau}}=0.
% \end{align}
% For successive derivatives, we rely on the general Leibniz rule to study the time-varying structure of the expression:
% \begin{align}\label{eq:xi_gen_leibniz_rule}
%     \frac{d^i\xi(t)}{dt^i}&=-2c\frac{d^{i-1}}{dt^{i-1}}\left(\mu_2(t,T)e^{-cT\left(\mu_2(t,T)-1\right)}\right)\nonumber\\
%     &=-2c\sum_{k=0}^{i-1}{i-1\choose k}\mu_2^{(k)}(t,T)\frac{d^{i-k-1}\xi(t)}{dt^{i-k-1}}\\
%     &=-2c\sum_{k=0}^{i-1}\frac{2^{\overline{k}}}{T^k}{i-1\choose k}\mu_{2+k}(t,T)\frac{d^{i-k-1}\xi(t)}{dt^{i-k-1}}.\nonumber
% \end{align}
% We assume by induction that~\eqref{eq:soft_landing_limit} holds for the $(i-1)$th derivative such that
% \begin{align}\label{eq:xi_induction_assumption}
%     \frac{d^{i-1}\xi(t)}{dt^{i-1}}&=\mathcal{P}^{2(i-1)}\left(\mu_1(t,T)\right)\xi(t);
% \end{align}
% it follows from applying l'H\^{o}pital's rule to~\eqref{eq:xi_gen_leibniz_rule} with~\eqref{eq:xi_induction_assumption} $i+2$ times that~\eqref{eq:soft_landing_limit} holds for all $i\in\mathbb{N}$.
% \end{proof}

We can now proceed with the proof of Theorem~\ref{thm:main}.%, where we select $t_0=0$ for clarity.

\begin{proof}
We first pursue non-positivity of $y(t)$ under~\eqref{ueqn},~\eqref{cieqn},~\eqref{ciueqn}. The system beginning from safety, that is, $y(t_0)=x_1(t_0)<0$, implies that $h_1(t_0)>0$.
We proceed by induction: suppose $h_i(t_0)>0$ for some $i=1,\dots,n-1$; it follows from~\eqref{hidoteqn} and differentiating~\eqref{hieqn} along~\eqref{sys} that
\begin{align}
%h_{i+1}(t_0) &= c_ih_i(t_0) + \dot h_i(t_0)\\
\hspace*{-0.113in}h_{i+1}(t_0) &= c_ih_i(t_0) - x_{i+1}(t_0) + \frac{d}{dt}\alpha_{i-1}(\underline{x}_{i-1}(t_0),t_0)
\end{align}
The \emph{initial} control gains~\eqref{cieqn},~\eqref{ciueqn} are designed so that
\begin{align}
    c_ih_i(t_0) - x_{i+1}(t_0) + \frac{d}{dt}\alpha_{i-1}(\underline{x}_{i-1}(t_0),t_0)>0,
\end{align}
where we've used~\eqref{hieqn}. With $h_i(t_0)>0$ $\forall i=1,\dots,n$, it is easy to show by backwards strong induction and the variation of constants formula that $h_i(t)>0$ for $t\in[t_0,t_0+T)$. %We now show that $h_i(t_0)>0$ for all $i=1,\dots,n$ is a sufficient condition for non-negativity of $h_i(t)$ for $t\in[t_0,t_0+T)$.

% REMOVED BEFORE FINAL SUBMISSION DUE TO SPACE LIMITATION
% Applying the Comparison lemma and variation of constants formula to~\eqref{hidoteqn},~\eqref{hndoteqn} gives
% \begin{align}
% h_i(t) &= h_i(t_0)e^{-c_i\int_{t_0}^t\mu_2(s)\,ds}\nonumber\\
% &\quad\quad + \int_{t_0}^t e^{-c_i\int_\tau^t\mu_2(s)\,ds}h_{i+1}(\tau)\,d\tau,\label{hisoln}\\
% h_n(t)&\geq h_n(t_0)e^{-c_n\int_{t_0}^t\mu_2(s)\,ds}>0,\label{hnineq}
% \end{align}
% for $t\in[t_0,t_0+T)$. Substituting~\eqref{hnineq} into~\eqref{hisoln} for $i=n-1$ yields
% \begin{align}
% h_{n-1}(t) &\geq h_{n-1}(t_0)e^{-c_{n-1}\int_{t_0}^t\mu_2(s)\,ds}\ + \nonumber\\
% &\quad\quad h_{n}(t_0)\int_{t_0}^t\Big[e^{-c_{n-1}\int_\tau^t\mu_2(s)\,ds}\nonumber\\
% &\quad\quad\quad\quad\quad\quad\quad \times e^{-c_n\int_{t_0}^\tau \mu_2(s)\,ds}\Big]\,d\tau\nonumber\\
% &\geq h_{n-1}(t_0)e^{-c_{n-1}\int_{t_0}^t\mu_2(s)\,ds}>0.\label{eq:h_n-1_inequality}
% \end{align}

% By using~\eqref{hnineq},~\eqref{eq:h_n-1_inequality} and by proceeding by backwards strong induction, it follows that
% \begin{align}
% h_1(t) \geq h_1(t_0)e^{-c_1\int_{t_0}^t\mu_2(s)\,ds}
% \end{align}
% which is equivalent to
% \begin{align}
% x_1(t) &\leq x_1(t_0)e^{-c_1\int_{t_0}^t\mu_2(s)\,ds}< 0.
% \end{align}
% for all $t\in[t_0,t_0+T)$.

We now pursue uniform boundedness of the the control law~\eqref{ueqn}. We partition the time horizon $[0,T)$ into intervals for which the system is either deemed safe or unsafe according to the safety filter \eqref{ueqn} by defining
\begin{align}
    \scalemath{0.93}{t_k:=\begin{cases}
    \min\{t_{k-1}<t\leq T:u_{\mathrm{nom}}(t)=\alpha_n(\underline{x}_n,t)\},&\text{if it exists},\\
    T,&\text{otherwise},
    \end{cases}}\nonumber
\end{align}
for $k\in\mathbb{N}$ with $t_0=0$, where
\begin{align}
    [0,T)=\bigcup_{\substack{k\in\mathbb{N}\cup\{0\} \\ t_{k+1}\leq T}} [t_k,t_{k+1}).\label{eq:time_partition}
\end{align}
We have constructed this partition such that the control law~\eqref{ueqn} remains continuous at $t_k$, precluding Zeno behavior of the closed-loop system. Since the system is initially safe, $t_1\neq T$ represents the first time that safety is enforced by~\eqref{ueqn}. 
For $t\in[t_{2k},t_{2k+1})$, $k\in\mathbb{N}\cup\{0\}$ and $t_{2k+1}\leq T$, we define % the CBFs
\begin{align}
h_i^{2k} &:= -x_i+\alpha_{i-1}^{2k}(\underline{x}_{i-1},t-t_{2k}),\label{hieqn2}\\
\alpha_i^{2k}(\underline{x}_i,t-t_{2k}) &:= c_{i}^{2k}\mu_2(t-t_{2k},T-t_{2k}) h_i^{2k}\nonumber\\
&\;+\frac{d}{dt}\alpha_{i-1}^{2k}(\underline{x}_{i-1},t-t_{2k})\label{alphaieqn2}
\end{align}
for $i=1,\dots,n$, with $\alpha_0^{2k}(\underline{x}_{0},t-t_{2k})\equiv 0$. 
% It follows from~\eqref{ueqn} that during these intervals, these CBFs satisfy
% \begin{align}
% \dot h_i^{2k} &= -c_i^{2k}\mu_2(t-t_{2k},T-t_{2k}) h_i^{2k} + h_{i+1}^{2k},\label{hidoteqn2}\\
% \dot h_n^{2k} &=-u_{\mathrm{nom}}+\frac{d}{dt}\alpha_{n-1}^{2k}(\underline{x}_{n-1},t-t_{2k}),\label{hndoteqn2}
% \end{align}
% for $i=1,\dots,n-1$. 
Similarly, for $t\in[t_{2k-1},t_{2k})$, $k\in\mathbb{N}$ and $t_{2k}\leq T$, we define the CBFs
\begin{align}
h_i^{2k-1} &:= -x_i+\alpha_{i-1}^{2k-1}(\underline{x}_{i-1},t-t_{2k-1}),\label{hieqn3}\\
\alpha_i^{2k-1}&(\underline{x}_i,t-t_{2k-1}) :=\frac{d}{dt}\alpha_{i-1}^{2k-1}(\underline{x}_{i-1},t-t_{2k-1})\label{alphaieqn3}\\
&\qquad+c_{i}^{2k-1}\mu_2(t-t_{2k-1},T-t_{2k-1})\nonumber h_i^{2k-1}
\end{align}
for $i=1,\dots,n$, with $\alpha_0^{2k-1}(\underline{x}_{0},t-t_{2k-1})\equiv 0$. We select $c_i^0=c_i$ according to~\eqref{cieqn},~\eqref{ciueqn}, and we select
\begin{align}\label{eq:c_i_k_choice}
    c_i^k=c_i^{k-1}\mu_2(t_{k}-t_{k-1},T-t_{k-1}),\quad k\in\mathbb{N}.
\end{align}
Since $\alpha_0^{2k}(\underline{x}_{0},t-t_{2k})=\alpha_0^{2k-1}(\underline{x}_{0},t-t_{2k-1})\equiv 0$, it follows that $h_1^{2k}(t_{2k-1})=h_1^{2k-1}(t_{2k-1})$ for $k\in\mathbb{N}$. Furthermore, by applying the initial gain selection~\eqref{eq:c_i_k_choice} to~\eqref{hieqn3},~\eqref{alphaieqn3} and comparing them to~\eqref{hieqn2},~\eqref{alphaieqn2} at $t=t_{2k-1}$, we deduce that $h_i^{2k}(t_{2k-1})=h_i^{2k-1}(t_{2k-1})$ for $i=2,\dots,n$. The same treatment leads to the equalities $h_i^{2k}(t_{2k})=h_i^{2k-1}(t_{2k})$ for $i=1,\dots,n$. Hence, the initial gain selection~\eqref{eq:c_i_k_choice} for each time partition in~\eqref{eq:time_partition} ensures that the system dynamics remain continuous at every time. In fact, it simply tracks the growth of the ``blow-up" function $\mu_2$ over the time intervals.

Furthermore, we can leverage Lemma~\ref{lemma:blow_ups_commute} and the initial gain selection~\eqref{eq:c_i_k_choice} to show that
\begin{align}
    \prod_{k\in\mathbb{N}} c_i^k\mu_2(t-t_k,T-t_k)=\mu_2(t,T);
\end{align}
in other words, the CBF design over the partitioned set~\eqref{eq:time_partition} is consistent with the design~\eqref{hieqn}--\eqref{ciueqn}.

For $t\in[t_{2k},t_{2k+1})$, $k\in\mathbb{N}\cup\{0\}$ and $t_{2k+1}\leq T$, the system is safe and the nominal control---which we assume to be uniformly bounded (continuous over a compact time interval)---is being applied. For
$t\in[t_{2k-1},t_{2k})$, $k\in\mathbb{N}$ and $t_{2k}\leq T$, we must estimate the size of the time-varying % control 
input to verify that it is bounded. It follows from~\eqref{ueqn} adapted as $u=\alpha_n^{2k-1}(\underline{x}_n,t-t_{2k-1})$ that during these intervals, the CBFs satisfy
\begin{align}
\dot h_i^{2k-1} &= -c_i^{2k-1}\mu_2(t-t_{2k-1},T-t_{2k-1}) h_i^{2k-1} + h_{i+1}^{2k-1},\label{hidoteqn3}\\
\dot h_n^{2k-1} &=-c_n^{2k-1}\mu_2(t-t_{2k-1},T-t_{2k-1})h_n^{2k-1},\label{hndoteqn3}
\end{align}
for $i=1,\dots,n-1$. To this end, we first study the stability of~\eqref{hidoteqn3},~\eqref{hndoteqn3}. We can solve~\eqref{hndoteqn3} explicitly to obtain
\begin{align}\label{eq:hn_2k-1_sol}
    &h_n^{2k-1}(t)\\
    &=e^{-c_n^{2k-1}(T-t_{2k-1})(\mu_1(t-t_{2k-1},T-t_{2k-1})-1)}h_n^{2k-1}(t_{2k-1})\nonumber,
\end{align}
whereas for $i=1,\dots,n-1$, we have the relationship
\begin{align}
    &h_i^{2k-1}(t)\nonumber\\
    &=e^{-c_i^{2k-1}(T-t_{2k-1})(\mu_1(t-t_{2k-1},T-t_{2k-1})-1)}h_i^{2k-1}(t_{2k-1})\nonumber\\
    &+\int_{t_{2k-1}}^t e^{-c_i^{2k-1}\int_{\tau}^t\mu_2(z-t_{2k-1},T-t_{2k-1})dz}h_{i+1}(\tau)d\tau.
\end{align}
We apply Lemma~\ref{lemma:exp_soft_landing} to~\eqref{eq:hn_2k-1_sol} to establish that successive derivatives of~\eqref{eq:hn_2k-1_sol} will converge to zero by the terminal time:
\begin{align}\label{eq:limit_der_h_n}
    \lim_{t\rightarrow T^{-}}\frac{d^r h_n^{2k-1}(t)}{dt^r}=0,\quad t\in[t_{2k-1}, T^-],
\end{align}
for $r\in\mathbb{N}\cup\{0\}$. For $i=1,\dots,n-1$, we compute
\begin{align}\label{eq:hi_2k-1_1st_der}
    &\frac{dh_i^{2k-1}(t)}{dt}=h_{i+1}(t) + h_i^{2k-1}(t_{2k-1})\times\\
    &\qquad\qquad\frac{d}{dt}\left(e^{-c_i^{2k-1}(T-t_{2k-1})(\mu_1(t-t_{2k-1},T-t_{2k-1})-1)}\right).\nonumber
\end{align}
By applying Lemma~\ref{lemma:exp_soft_landing} to the second term within~\eqref{eq:hi_2k-1_1st_der}, and by backward strong induction on~\eqref{eq:limit_der_h_n}, we get for $r\in\mathbb{N}\cup\{0\}$% that
\begin{align}\label{eq:limit_der_h_i}
    \lim_{t\rightarrow T^{-}}\frac{d^r h_i^{2k-1}(t)}{dt^r}=0,\quad i=1,\dots,n-1.
\end{align}

\begin{figure*}[!t]
\centering
\includegraphics[clip,width=.9\textwidth]{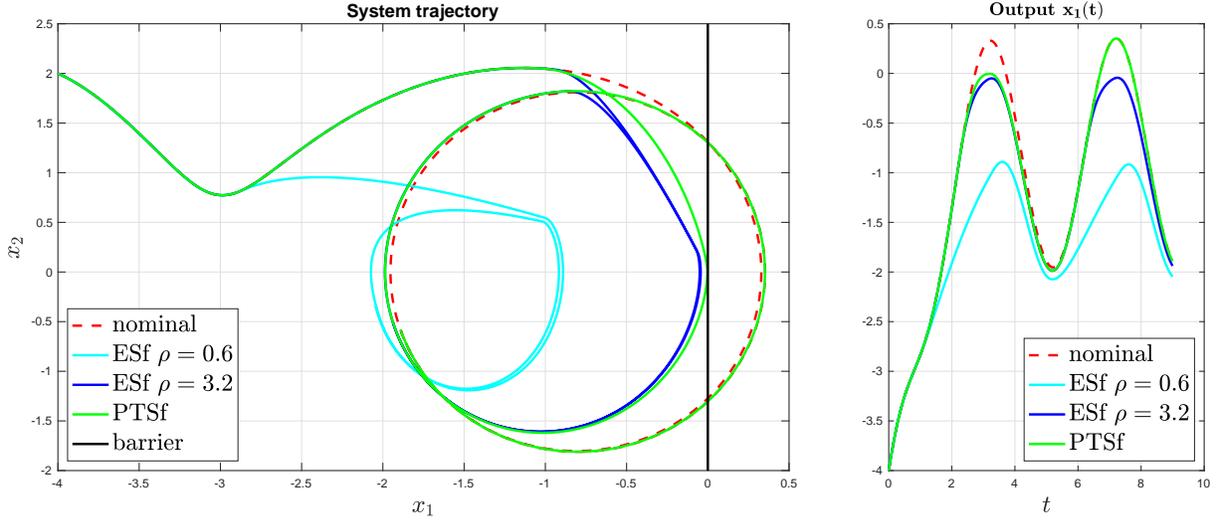}
\caption{System trajectories (left) and outputs (right) for double integrator under nominal controller \eqref{eq:exampleunom} with terminal time $T=4$, and initial condition $(x_1(0),x_2(0))=(-4,2)$. The PTSf safety-filter uses \eqref{ueqn} with $c_2=c_1=0.6$ while the ESf safety-filter uses \eqref{eq:safety_filter_ti} with $\rho=0.6$ and $\rho=3.2$ -- the latter value tuned to make ESf react at the same instant as PTSf.}
\label{fig:figure1}
\end{figure*} 

\begin{figure*}[!t]
\centering
\includegraphics[clip,width=.9\textwidth]{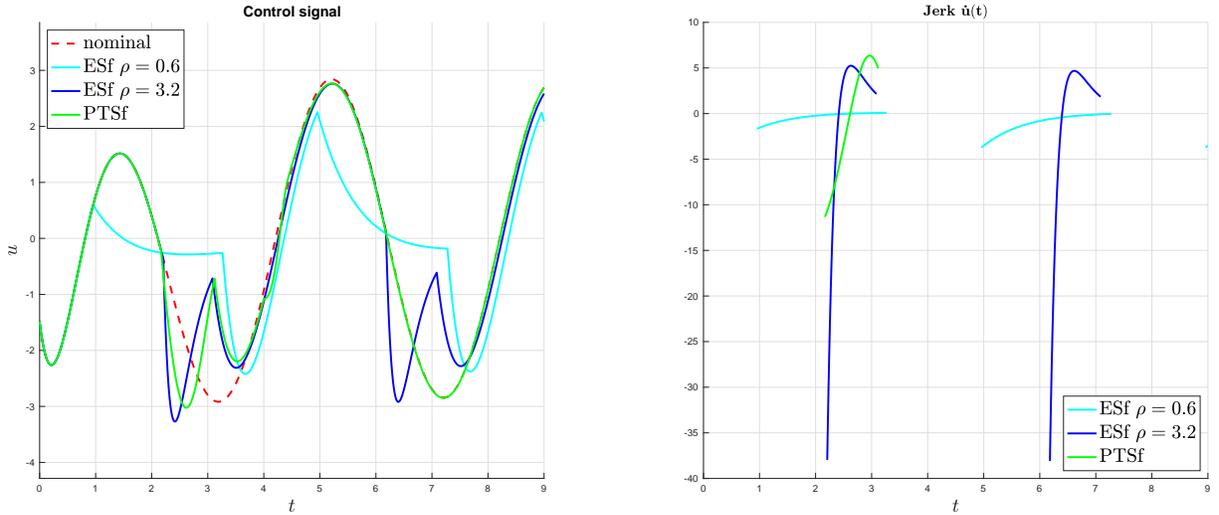}
\caption{Left: Control signal. Right: Jerk during intervals when nominal command is overridden. When ESf is tuned ($\rho=3.2$ case) to be less conservative like PTSf, the magnitude of the jerk increases significantly.}
\label{fig:figure2}
\end{figure*} 

Hence, we have shown that when the nominal controller is overridden by the safety filter to enforce safety during $t\in[t_{2k-1},t_{2k})$, $k\in\mathbb{N}$ and $t_{2k}\leq T$, our time-varying backstepping design ensures that the CBF converge very smoothly to zero by the terminal time---indeed, all of their derivatives also converge to zero by the terminal time.

By using~\eqref{hidoteqn3} within the derivative term of~\eqref{alphaieqn3} in an iterative fashion, we can verify by induction that~\eqref{alphaieqn3} for $i=n$ is equivalent to
\begin{align}\label{eq:PTS_controller_new_form}
    \alpha_n^{2k-1}&(\underline{x}_n,t-t_{2k-1}) =\sum_{r=1}^{n-1}\frac{d^r}{dt^r}\left(h_{n-r}^{2k-1}-\dot{h}_{n-r-1}^{2k-1}\right)\\
&\qquad+c_{n}^{2k-1}\mu_2(t-t_{2k-1},T-t_{2k-1})\nonumber h_n^{2k-1}.
\end{align}
We conclude from~\eqref{eq:limit_der_h_n},~\eqref{eq:limit_der_h_i} and applying l'H\^{o}pital's rule to~\eqref{eq:PTS_controller_new_form} as before that
\begin{align}
    \left|\alpha_n^{2k-1}(\underline{x}_n,t-t_{2k-1})\right|<+\infty,\quad\forall\ t\in[t_{2k-1},t_{2k}),
\end{align}
for $k\in\mathbb{N}$ and $t_{2k}\leq T$. %This concludes our proof of controller uniform boundedness.
\end{proof}

\section{Double Integrator Design Interpretation}\label{sec:sim}
%\subsection{PTSf versus ESf: The Peaking Phenomenon}
We consider the double integrator i.e. system \eqref{sys} with $n=2$. Suppose that the nominal control input $u_{\mathrm{nom}}$ is at risk of making the system unsafe, and we wish to design a \emph{time-invariant} safety filter that overrides the nominal controller and takes the system to the origin. This problem was studied in~\cite[Sec. 3.B]{nguyen2016exponential} for input-output linearized systems via pole-placement (which inherently relies on the backstepping method---see~\cite[Rk. 5]{nguyen2016exponential}), which achieves exponential convergence to the origin with arbitrary decay rate. We define the barrier functions $h_1:=-x_1$, $h_2:=-x_2+\rho h_1$, $\rho>0$, % \begin{equation}
%     \begin{aligned}
%     h_1&:=-x_1,\\
%     h_2&:=-x_2+\rho h_1,\quad \rho>0,
% \end{aligned}\label{EShtransform}
% \end{equation} 
with the goal of keeping $h_1\geq 0$ uniformly. Consider the following time-invariant safety filter designed as in~\cite{nguyen2016exponential}:
\begin{align}
u &= \min\left\{u_{\text{nom}},-\left(2\rho^2\;\; 3\rho\right)x \right\},&\text{for }t_0\leq t<\infty,\label{eq:safety_filter_ti}
\end{align}
with $\rho\geq\max\left\{0,-x_2(t_0)/x_1(t_0)\right\}$. Suppose the safety-filter overrides $u_{\text{nom}}$ at $t=t_0+\bar{t}<\infty$ and continues to enforce safety thereafter (i.e., $u(t)=-\left(2\rho^2\;\; 3\rho\right)x(t)$ for all $t\geq t_0+\bar{t}$, placing the closed-loop poles for the $x$-system at $\{-\rho,-2\rho\}$). Then the closed-loop system is given by
\begin{align}\nonumber\\[-0.2in]\label{eq:cl_solution_static_ti}
    &x(t)=e^{-\rho(t-t_0-\bar{t})}\\
    &\times\left(\begin{array}{c c} 2-e^{-\rho(t-t_0)}&\frac{1}{\rho}-\frac{e^{-\rho(t-t_0-\bar{t})}}{\rho}\\2\rho\left(e^{-\rho(t-t_0-\bar{t})}-1\right)&2e^{-\rho(t-t_0-\bar{t})}-1\end{array}\right)x(t_0+\bar{t}).\nonumber
\end{align}
If we wish to achieve large exponential decay when the system is unsafe, we can select $\rho\gg \max\left\{0,-x_2(t_0)/x_1(t_0)\right\}$ as large as desired. However, for small $t-t_0-\bar{t}$, the righthand side of~\eqref{eq:cl_solution_static_ti} can be very large depending on the size of $\rho$ (in particular, $x_2$ grows with $\rho$). This illustrates the ``peaking" phenomenon, which was studied for ODE control systems in~\cite{kimura1981new,sussmann1991peaking,khalil1992semiglobal}. %The celebrated work of Sussmann and Kokotovic~\cite{sussmann1991peaking} has exposed the possibility of disastrous outcomes in input-output feedback linearization, where rapid regulation of the output can have catastrophic consequences on the zero dynamics. Moreover, due to the structure of the feedback~\eqref{eq:safety_filter_ti}, even for systems with a full relative degree (systems without zero dynamics, as above), the control input becomes extremely large near time $t_0+\bar{t}$.
% In the context of safety, if $(x_1,x_2)$ represent position and velocity, seeking time-invariant safety filters with large exponential decay ($\rho\gg\max\left\{0,-\frac{x_2(t_0)}{x_1(t_0)}\right\}$) results in a very large and rapid transient response in the velocity, which is undesirable as it causes a large ``jerk" to the system. For our time-varying safety-filter~\eqref{eq:safety_filter_ti}, the control gains are chosen to initially start quite small and depending on the initial conditions (see~\eqref{eq:safety_override_tv}), and only grow very large simultaneously as the states grow very small and as time approaches the prescribed terminal time. This eliminates the possibility of peaking.
We now compare these results graphically to demonstrate the advantages of time-varying backstepping. We perform numerical simulations for the double-integrator system under the nominal controller 
\begin{equation}
u_{\text{nom}}=-4\big[x_1 + \sin(\omega t) + 0.8\big] - 4\big[x_2 + \omega\cos(\omega t)\big]\label{eq:exampleunom}
\end{equation}
with $\omega=2\pi/T$ where $T=4$ is the prescribed time. For initial condition $x(0)=(-4,2)^\top$, we use the time-varying PTSf safety-filter \eqref{ueqn} with choice of gains $c_2=c_1=\max\left\{0,-x_2(0)/x_1(0)\right\}+0.1=0.6$ and use ramp function \eqref{eq:rampfunction} with $m=2$ and $\bar T=0.5$. % for controller continuity at $t_0+T$ (see Section \ref{sec:nthorder}, Eq. \eqref{eq:rampfunction} for details).
For comparison, we use the time-invariant ESf safety-filter \eqref{eq:safety_filter_ti} with $\rho=0.6$ and $\rho=3.2$. The choice $\rho=0.6$ was made to allow a gain equivalent to the initial gains of PTSf, and the choice $\rho=3.2$ was tuned to make ESf less conservative and to react at around the same instant as PTSf. For numerical stability near the origin, we clip the blow-up function $\mu_2$ at a maximum value $\mu_{2,\max}=1000$ --- which still allows the PTSf gains grow to several orders of magnitudes larger than $\rho=3.2$. The system trajectories under PTSf and ESf are shown in Fig.~\ref{fig:figure1}. %where we observe the ESf filter with $\rho=0.6$ being overly conservative, overriding the nominal trajectory much sooner, despite being significantly far from the barrier. With $\rho$ increased to $3.2$, the ESf filter becomes less conservative like PTSf but at the cost of a significantly higher jerk as evident in Fig.~\ref{fig:figure2}. %shows the resulting control signal and jerk in the time interval when the safety overrides are active. %Here, in the $\rho=3.2$ case where the ESf filter has been tuned to intervene at the same time as the PTSf filter, the resulting jerk becomes more significant. 
There, we observe that while ESf can be tuned to be less conservative like PTSf by choosing larger gains, it comes at the expense of a significant jerk. %Lastly, as evident in both figures, the PTSf filter eventually allows the system evolve freely after the prescribed time $T=4$ has elapsed.

%We now present our prescribed-time safety-filter design for the $n$th order chain-of-integrator systems. While the time-varying backstepping technique is applicable to a wider class of systems (\emph{nonlinear} strict feedback ones), we opt for increased clarity by limiting our exposition to linear systems.

% \section{Future Work}
% In this work, we present a safety filter design for a chain of integrators that enforces an output constraint for a prescribed time. Our design uses a time-varying backstepping transformation with gains that grow as time approaches the terminal time. Despite the use of gains that grow towards infinity, we show that our safegaurding controller remains uniformly bounded provided that the nominal controller is uniformly bounded. While we have limited our study to a chain of integrators, the treatment readily extends to nonlinear strict feedback systems. 
% Compelling future research directions include: %studying predictor-based safety-filter designs to compensate for input delays; 
% extending our time-varying backstepping treatment to higher-dimensional nonlinear constraints; characterizing and compensating for the effect of persistent disturbances appearing on the right-hand side of the dynamics; and developing a discretization algorithm that preserves the properties of our prescribed-time safety-filter.

\bibliographystyle{IEEEtranS}
\bibliography{paper}

\end{document}